\documentclass{sigchi-ext}
\usepackage[T1]{fontenc}
\usepackage{textcomp}
\usepackage[scaled=.92]{helvet} 
\usepackage{graphicx} 
\usepackage{balance}  
\usepackage{booktabs} 
\usepackage{ccicons}  
\usepackage{ragged2e} 

\def\plaintitle{SIGCHI Extended Abstracts Sample File: Note Initial
  Caps} 
\def\emptyauthor{}
\def\plainkeywords{Authors' choice; of terms; separated; by
  semicolons; include commas, within terms only; required.}

\title{ACE, Action and Control via Explanations: A Proposal for LLMs to Provide Human-Centered Explainability for Multimodal AI Assistants}

\numberofauthors{6}
\author{%
  \alignauthor{%
    \textbf{Elizabeth Anne Watkins}\\
    \affaddr{Intel Labs} \\
    \affaddr{Santa Clara, CA 95054, USA} \\
    \email{Elizabeth.Watkins@intel.com} }\alignauthor{%
    \textbf{Emanuel Moss}\\
    \affaddr{Intel Labs} \\
    \affaddr{Santa Clara, CA 95054, USA} \\
    \email{Emanuel.Moss@intel.com} } \vfil \alignauthor{%
    \textbf{Ramesh Manuvinakurike}\\
    \affaddr{Intel Labs}\\
    \affaddr{Hillsboro, OR 97124, USA}\\
    \email{Ramesh.Manuvinakurike
    @intel.com} }\alignauthor{%
    \textbf{Meng Shi}\\
    \affaddr{Intel Labs}\\
    \affaddr{Santa Clara, CA 95054, USA}\\
    \email{Meng.Shi@intel.com} } \vfil \alignauthor{%
    \textbf{Richard Beckwith}\\
    \affaddr{Intel Labs}\\
    \affaddr{Hillsboro, OR 97124, USA}\\
    \email{Richard.Beckwith@intel.com} \\
    }
    \alignauthor{%
    \textbf{Giuseppe Raffa}\\
    \affaddr{Intel Labs}\\
    \affaddr{San Diego, CA 92109, USA}\\
    \email{Giuseppe.Raffa@intel.com} } }

\definecolor{linkColor}{RGB}{6,125,233}
\hypersetup{%
  pdftitle={\plaintitle},
  pdfauthor={\emptyauthor},
  pdfkeywords={\plainkeywords},
  bookmarksnumbered,
  pdfstartview={FitH},
  colorlinks,
  citecolor=black,
  filecolor=black,
  linkcolor=black,
  urlcolor=linkColor,
  breaklinks=true,
}

\begin{document}

\CopyrightYear{2024}
\setcopyright{rightsretained}
\conferenceinfo{CHI Workshop on Human-Centered Explainable AI '24,}{May 2024, HI, USA}
\isbn{978-1-4503-6819-3/20/04}
\doi{https://doi.org/10.1145/3334480.XXXXXXX}
\copyrightinfo{\acmcopyright}

\maketitle

\RaggedRight{} 

\begin{abstract}
  In this short paper we address issues related to building multimodal AI systems for human performance support in manufacturing domains. We make two contributions: we first identify challenges of participatory design and training of such systems, and secondly, to address such challenges, we propose the ACE paradigm: "Action and Control via Explanations". Specifically, we suggest that LLMs can be used to produce explanations in the form of human interpretable "semantic frames", which in turn enable end users to provide data the AI system needs to align its multimodal models and representations, including computer vision, automatic speech recognition, and document inputs. ACE, by using LLMs to "explain" using semantic frames, will help the human and the AI system to collaborate, together building a more accurate model of humans activities and behaviors, and ultimately more accurate predictive outputs for better task support, and better outcomes for human users performing manual tasks.
\end{abstract}



\section{Introduction}
This short position paper proposes a research
agenda examining how LLMs intersect with explainability in the participatory design of a multimodal AI system to support task performance in manufacturing, a highly specific and specialized system, domain, and context.  We examine what kinds of problems can be addressed by asking an LLM to "explain" the system, and 
propose that to best support both end-users and developers who are collaborating to build this system, explanations must go beyond developer-focused technical information about how the system operates, and toward an end-user focused, "mechanistic" or "functional" explanation \cite{liao2023ai} that supports humans' practical, goal-oriented use of a system.

In this position paper, we first describe the industrial setting in which this development is taking place and a challenge we have observed in human-in-the-loop, participatory training of a multimodal AI system. We then describe findings in prior research and our empirical observations of end-users/collaborators as they interacted with an AI assistant, 
and described what explanatory information they would like to receive 
as they provided training to the system. We apply insights drawn from these observations to the era of LLMs. 

Specifically, in the context of multimodal learning for performance support, this proposal suggests prompting users, via explanations, to deploy 
a specific linguistic conversational format of ``semantic frames." This format, we suggest, better supports human interaction with -- and training of -- the AI assistant. Ultimately,  this may be a more useful form of explanation than commonplace wisdom in XAI, which tends to focus on providing system explanations,  might otherwise suggest. We term this proposed system ACE: Action and Control via Explanations. 

\section{XAI and Human-AI Collaboration}

Explanations are highly context-specific. They are units of information that cannot stand on their own but are given meaning through the interactions between user and system (and by extension, developers), and the context in which that interaction takes place~\cite{dourish2004we}.  Their effectiveness depends on \textit{what} is being explained \textit{to whom} \cite{ehsan2021explainable} for \textit{what purpose}.

Prior work on explainable AI has identified that different stakeholders have different "interests, goals, expectations, needs, and demands" \cite{langer2021we} for AI systems. End users who are not AI experts have described that their interests are not in better "understanding" precisely how a system functions. Rather, end users ask for information that can help them become better "collaborators" with these systems, i.e. for guidance around how to change their behaviors to provide better inputs to AI systems in order to obtain better or more useful, more accurate outputs \cite{kim2023help}. 
As AI systems come to play a larger role in people's lives, it is critically important that human-centered explanations support human-AI interaction and enable successful outcomes across deployment domains. 

\section{Setting}


This project centers around an in-development system called MARIE (\textbf{M}ultimodal \textbf{A}ctivity \textbf{R}ecognition in an \textbf{I}ndustrial \textbf{E}nvironment). This first implementation of MARIE is intended to provide performance support to end users who  perform intricate cleaning tasks inside our manufacturing facilities. 
The stakes of this work are high, as any debris left on manufacturing components can have detrimental and costly consequences. MARIE is intended to help technicians ensure their cleaning processes are complete, to reduce losses due to debris on parts. MARIE recognizes where a technician is in their process and offers tips, reminders, and guidance, supporting training for novices and providing support for experts. 

MARIE's development process is intended to address a known bottle-neck in AI development: data representing this kind of proprietary, highly specific information is difficult to collect and label. MARIE's solution is to use participatory design and engage domain-expert end-users as collaborators. These experts in the cleaning process provide crucial data to ``teach" MARIE about what they do via demonstration and concurrent conversational exchanges with the system.  
Their voice is captured by microphones and processed using automatic speech recognition (ASR), and the visual context is captured by cameras and processed using computer vision (CV) models such as action recognition and object recognition. These inputs are combined with previously analyzed in-domain process information, e.g., "spec" documents containing step-by-step instructions. 
MARIE brings together all of these inputs to attain understanding of the cleaning process and provide performance support to both novices for training, as well as expert technicians as they juggle tasks in the facility. 

\section{Contributions: A Challenge of Multimodal 
\\Systems, and a Proposal for Structuring LLM 
\\Explanations}

\begin{marginfigure}[-5pc]
  \begin{minipage}{\marginparwidth}
    \centering
    \includegraphics[width=0.9\marginparwidth]{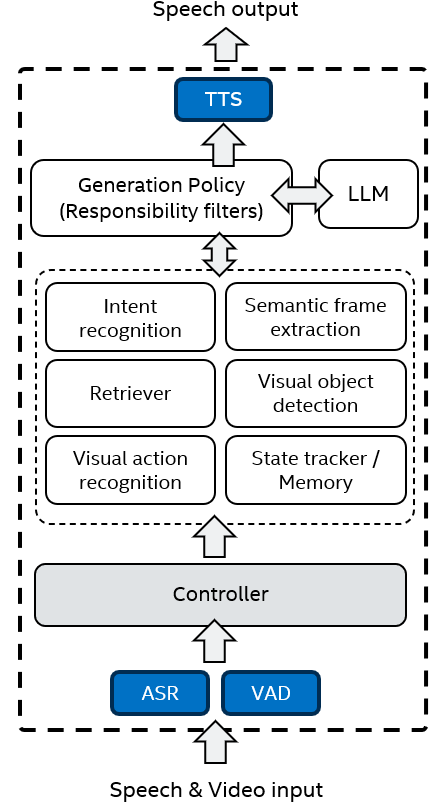}
    \caption{Shows the LLM-based architecture of the system. The system consumes inputs from multimodal }~\label{fig:marginfig}
  \end{minipage}
\end{marginfigure}


\textbf{A Challenge of Multimodal Systems} A key component of MARIE's ``training" process, and what will enable MARIE to understand human actions well enough to provide accurate tips and guidance for performance support, is that MARIE must be able to align predictions made using disparate modes of information input. These inputs include vision data, audio data, and written data. Here a  challenge emerges: MARIE must ``understand" what activity (as ``seen" by the vision models, including action recognition and object recognition) is being performed, and align these vision model predictions with ASR predictions and data from the spec. This challenge can be observed in several mismatches between the naturalistic way that humans train MARIE through conversational exchange, and the information that MARIE needs to appropriately infer actions from multimodal data:


\textit{Missing information/clarification:} technicians commonly describe their actions in ways that are missing context --- context which MARIE needs. For example, an utterance like ``I'm cleaning this part" 
is missing key information about tools. Not only does the generic noun fail to specifcy the part being cleaned but also there is no indication of what the part is being cleaned with. Semantic parsing of the utterance can tell us that the ``action" being performed is  (clean) but not the item affected by the action-- the ``receiver", we know only (part). 


\textit{Misalignment:} 
there is typically temporal misalignment between speech and action, due to the fluidity with which humans tend to plan, think, and speak about their own behaviors. The actions that technicians physically perform, and the actions they describe themselves doing, are rarely synchronously aligned. For example, technicians often describe that they are ``about to use the brush" while they are still physically conducting the preceding step. While humans can easily understand conducting a current action while describing a future action, this produces misalignment and misunderstanding for the multimodal MARIE.


\textit{Alternative process execution:} The spec document often gives freedom to technicians to execute cleaning in different sequences. Certain actions might be performed at any point and a strict process execution order is not imposed. However, there remain certain actions that require strict execution after satisfying certain pre-conditions. Learning these alternative processes remains a challenge. 


\textbf{Semantic Frames for Actionable LLM Explanations}
In early empirical investigations of how technicians perceive the utility of explanations in MARIE, they expressed a desire for information that could help them to craft inputs to MARIE which would help MARIE to ``learn better." 
In our prior research in this space we proposed a prototype and showed the applicability of semantic frames for task guidance\cite{manuvinakurike2022human}. The current contributions builds on this work and draws on the field of human-centered explainability, to propose that explanations be deployed using ``semantic frames" structure, to enable humans to act in the most productive way to train MARIE. Using ``semantic frames" means that an LLM-based conversational agent would ``explain" MARIE's outputs by asking the end-user for needed data at the same time - using the ``semantic frame" as a container of meaning, which MARIE needs the human to fill. This could look, for example, like explaining why a predicted tip is incorrect, in the form of explicitly telling human users which components of a conversational input are needed. In this way, the explanation becomes ``actionable" by guiding humans to input a non-naturalistic form of language which better supplies the information the AI system needs for it to align its many modules to reach sufficient ``understanding." These 
linguistic inputs help the human and the AI system to collaborate, together building a more accurate model of human actions, and ultimately more accurate inferences, better task support, and better outcomes for human users.

\begin{marginfigure}[-5pc]
  \begin{minipage}{\marginparwidth}
    \centering
    \includegraphics[width=0.9\marginparwidth]{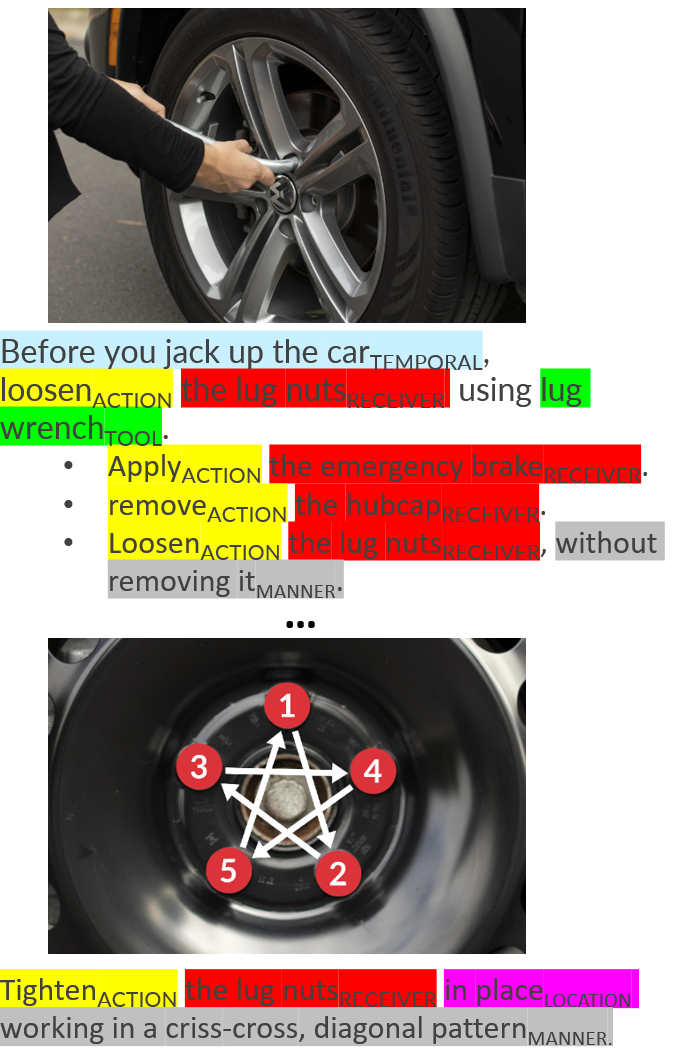}
    \caption{Semantic frames captured from task guidance scenarios.}~\label{fig:marginfig}
  \end{minipage}
\end{marginfigure}

Our system is multimodal, yet there is no native joining of vision and sound into a unified channel. We use language about the visual channel to link the modalities, which can be referred to as the symbol grounding problem \cite{harnad1990symbol}. The semantic frames we propose are abstractions of natural language  \cite{fillmore2006frame}, which ensure that a conversation includes specific components. These semantic frames can then be used for task guidance (e.g., question-answering, helping with tools, informing the extent of action, etc.) by including these as a part of dialogue system architecture. We create an ontology of actions with a semantic frame \cite{fillmore2001frame} to represent each action. Each semantic frame has both an action name “head word” and a number of roles related to that specific action (e.g., tool, location, or purpose). In our setting, these components consist of an action, and corresponding entities that help accomplish the action, or indicate temporal aspects of the action (e.g., [Action: Turn] [Receiver: the nut] using [Tool: a screw driver] [Extent: until snug]). Starting with relatively empty frames (one for each expected action), task experts fill the roles based on verbal interaction as they do the task and these are temporally aligned with actions to ground the reference.  Grounding the reference in interaction with experts unifies sound and vision into a single stream. Completed semantic frames allow the system to infer which action is referred to by an utterance.


For this participatory, human-in-the-loop training project, in which human technicians ``teach" an LLM system to recognize what they are doing on the factory floor, explanations must explain moments in which the system has incomplete semantic grounding infer an action or object, i.e., where the technician is in the protocol, or what object she holds. The technician can provide additional information that will complete the ``semantic frame" in which the task is occurring. Semantic frames can infer ``obvious" missing info that can be inferred logically depending on the process: For instance, ``pushing swabs now" means they're cleaning a specific item ("pushing" can only be done for the one item which is hollow). 

We can consider an illustrative example of how this might work in practice. For the purposes of this example, we can follow the procedure that's described in Figure 2, the act of changing a tire. (We've chosen to use this as our example, rather than provide details of the procedure that MARIE was initially designed for, in order to avoid the possibility of inadvertently disclosing protected IP.) Let's say that someone is beginning to change a tire. They have set up MARIE to get tips and reminders, and hold up a car jack and say to MARIE, "I am going to use this on the car." This spoken description lacks referent terms - the sentence doesn't provide MARIE with key data about what "this" means - the object that's being held, or which step the person is on in their task. At this point, MARIE makes a mistake. The system might mistake the step that the person is currently performing and guide them to the incorrect "next" step; the system might mistake what the person is holding and erroneously provide tips on how to handle the hallucinated object - for example, recommending that the person turn the lug wrench, while they are still holding the car jack. Noticing the error, the person then reasonably asks for an explanation in real time (as we have seen in empirical observations when MARIE has an error in its output, or the technician disagrees with MARIE's recommendation) - "no, that's not the next step, why did you think it was?" or "no, that's not what I'm holding. Why did you say that?" MARIE then provides an LLM-generated response based on the combined best available predictions from each of the contributing modules - with the semantic frame containing the referent objects that help MARIE reach "understanding" across modalities. In our example, MARIE might say "I see that you are holding a lug wrench, so I recommended turning the lug wrench to loosen the lug nuts." The intended effect of this explanation is to 
prompt the person with a sentence containing the erroneous referent terms, so that they can respond with the appropriate corrections, i.e. "I am not holding a lug wrench, I am holding a jack that I will use to jack up the car." The power of the semantic frame is in its application to elicit and sync referring and referent terms. These terms can be used to ground visual and language modalities, providing a set of references that can be found across models of object recognition, action recognition, and language. This grounding increases the probability that MARIE will accurately understand what object a person is holding and what steps come next.

Some limitations of this structure relate to cognitive overload or burden: providing too much information in an explanation can  impede or overwhelm decision-making. Too much of a cognitive load from an AI explanation can damage a user's confidence in the system \cite{hudon2021explainable}. The mitigation strategy of semantic frames, as a component of ACE, or Action and Control via Explantions, is to deliberately design explanations to integrate smoothly with user's workflow: users are provided with the explanation only when they ask for it, and the explanation itself - structured as a semantic frame - is focused on information which enables user action.  

The motivation behind ACE is that explanations could be more useful, both for humans being asked to provide data to ultimately achieve task performance support, and for developers building this system, if the inputs are "semantic frames." Although the semantic frame may sacrifice a little bit of naturalness of conversation, it provides transparency into the system's intelligence and reduces the user's cognitive load and improves system performance, as it clearly conveys the system's understanding and reasoning \cite{murad2023s}. So we propose semantic frames as a useful container for deriving meaning and aligning a multimodal system around a "ground truth."

\balance{} 

\bibliographystyle{SIGCHI-Reference-Format}
\bibliography{sample}

\end{document}